\documentclass[10pt,twocolumn]{IEEEtran}
\makeatletter
\def\subsubsection{\@startsection{subsubsection}
                                 {3}
                                 {\z@}
                                 {0ex plus 0.1ex minus 0.1ex}
                                 {0ex}
                             {\normalfont\normalsize\bfseries}}
\makeatother
\usepackage[T1]{fontenc}
\usepackage{ulem}
\usepackage{amsmath,enumitem}
\allowdisplaybreaks
\usepackage{hhline}
\usepackage[dvips]{graphicx}
\usepackage{yfonts,color}
\usepackage{soul,xcolor}
\usepackage{verbatim}
\usepackage{flushend}
\usepackage{amsmath}
\allowdisplaybreaks
\usepackage{amssymb}
\usepackage{amsthm}
\usepackage{float}
\usepackage{bm}
\usepackage{url}
\usepackage{array}
\usepackage{cite}
\usepackage{tikz}
\usepackage{framed}
\usepackage{balance}
\usepackage{hyperref}
\usepackage{booktabs}
\usepackage{courier}
\usepackage{pseudocode}
\usepackage{enumerate}
\usepackage{caption}
\usepackage{subcaption}
\usepackage{algorithm}
\usepackage[noend]{algpseudocode}

\newcommand{\rom}[1]{\uppercase\expandafter{\romannumeral #1\relax}}
\usepackage{color}
\usepackage{soul,xcolor}

\DeclareMathOperator*{\argmax}{arg\,max}
\DeclareMathOperator*{\argmin}{arg\,min}
\usepackage{cancel}
\usepackage{setspace}
\usepackage{glossaries}
\newacronym{tx}{Tx}{transmitter}
\newacronym{rx}{Rx}{receiver}

\title{Multiscale Adaptive Scheduling and Path-Planning for Power-Constrained UAV-Relays via SMDPs}
\author{Bharath Keshavamurthy~\IEEEmembership{Student Member, IEEE} and Nicol\`{o} Michelusi~\IEEEmembership{Senior Member, IEEE}
\thanks{An extension to this work has been submitted to IEEE TCCN \cite{TCCN_UAV}.}
\thanks{This work has been supported by NSF under grant CNS-2129015.}
\thanks{The authors are with Electrical, Computer and Energy Engineering,}
\thanks{Arizona State University. Email: \{bkeshav1, nicolo.michelusi\}@asu.edu.}
\vspace{-10mm}}

\begin{document}
\bstctlcite{IEEEexample:BSTcontrol}
\maketitle
\thispagestyle{empty}
\pagestyle{empty}
\setulcolor{red}
\setul{red}{2pt}
\setstcolor{red}
\newcommand{\linespreadexceptabstractandindex}{\setstretch{0.98}}
\newcommand{\linespreadforabstractandindex}{\setstretch{0.991}}
\newcommand{\extraspacebeforesec}{-4mm}
\newcommand{\extraspacebeforesubsubsec}{0.2mm}
\linespreadexceptabstractandindex

\begin{abstract}
We describe the orchestration of a decentralized swarm of rotary-wing UAV-relays, augmenting the coverage and service capabilities of a terrestrial base station. Our goal is to minimize the time-average service latencies involved in handling transmission requests from ground users under Poisson arrivals, subject to an average UAV power constraint. Equipped with rate adaptation to efficiently leverage air-to-ground stochastics, we first derive the optimal control policy for a single relay via a semi-Markov decision process formulation, with competitive swarm optimization for UAV trajectory design. Accordingly, we detail a multiscale decomposition of this construction: outer decisions on radial wait velocities and end positions optimize the expected long-term delay-power trade-off; consequently, inner decisions on angular wait velocities, service schedules, and UAV trajectories greedily minimize the instantaneous delay-power costs. Next, generalizing to UAV swarms via replication and consensus-driven command-and-control, this policy is embedded with spread maximization and conflict resolution heuristics. We demonstrate that our framework offers superior performance vis-à-vis average service latencies and average per-UAV power consumption: $11\times$ faster data payload delivery relative to static UAV-relay deployments and $2\times$ faster than a deep-Q network solution; remarkably, $1$ relay with our scheme outclasses $3$ relays under a joint successive convex approximation policy by $62$\%.
\end{abstract}


\glsresetall

\vspace{-4mm}
\section{Introduction}\label{I}
With sustained device proliferation, enterprises across sectors have stepped-up their adoption of Unmanned Aerial Vehicles (UAVs) to gather data, survey infrastructure, monitor operations, and automate logistics \cite{UAVSurvey}. Inevitably, this has fostered varied academic research and industrial R\&D on drone-augmented beyond line-of-sight connectivity and traffic offloading in cellular networks: the coverage and service capabilities of an extant terrestrial radio access network are enhanced by the mobility and maneuverability of these autonomous aerial relays \cite{FundamentalTradeoffs}. Unsurprisingly, the pervasive potential of such hybrid networks brings along a plethora of challenges in real-world deployments \cite{FundamentalTradeoffs}: specifically, on-board energy constraints of these aerial platforms impacting mission times, stringent Quality-of-Service (QoS) mandates for reliable connectivity, channel characteristics of Air-to-Ground (A2G) links in highly-mobile settings, and computational feasibility challenges in trajectory design brought on by the inherently large state and action spaces. Ergo, several works in the state-of-the-art have tried to tackle these challenges using tools from optimization theory, machine learning, and reinforcement learning---however, various problems remain unsolved and various challenges are left unaddressed.

\noindent{\textbf{Related Work}}: Perusing single UAV-relay formulations in current literature, we observe non-adaptive schemes \cite{SCA, PAoI, Rician} designed for applications where the IoT devices possess local storage or aggregation capabilities allowing for deterministic arrivals of data packets. Yet, practical deployments involve dynamically-generated traffic from miscellaneous sets of users, each with varying degrees of QoS mandates and technological prowess. Unlike these works, we consider dynamic traffic generation from random deployments of ground users, thereby constructing a control strategy that is receptive to uncertain system dynamics. Furthermore, these works solve for the optimal service schedules and associated trajectories via Successive Convex Approximation (SCA) \cite{SCA, PAoI, Rician}, which apart from being computationally infeasible to accommodate dynamic traffic due to prohibitively large convergence times, relies on first-order Taylor approximations of the optimization problem to enforce convexity, thereby introducing inaccuracies into the model. On another note, these works employ Free Space Path-Loss (FSPL) models that fail to account for the A2G channel characteristics inherent in UAV-assisted wireless networks; moreover, their approximations in the traffic delivery constraint preclude the adoption of rate adaptation which allows the transmitters in the network to leverage channel stochastics to maximize throughput. In this paper, in addition to accurately modeling A2G channel characteristics and employing rate adaptation at all the transmitters to efficiently exploit said characteristics, there are no such underlying approximations.

Pivoting to the path-planning problem for a single relay, a Competitive Swarm Optimization (CSO) \cite{CSO} approach is proposed in this paper to bypass the computational infeasibility seen in \cite{SCA, PAoI, Rician}. Unlike SCA, which employs approximations to enforce convexity, CSO does not depend on the specific problem structure to work effectively. Contrary to the limited update scope of Particle Swarm Optimization (PSO) \cite{PSO}, CSO exhibits superior performance on large-scale optimization benchmarks \cite{CSO}, since it involves more efficient updates wherein pair-wise competition is invoked between particles---permitting the winners to advance and the loser particles to learn from the winners. Unreasonably, works that employ PSO, either optimize static hovering positions only \cite{Efficient3DPlacementPSO}, or impose impractical path and velocity restrictions \cite{PSOPathStructure, PAoI}.

Next, shifting our attention to swarm orchestration frameworks, we find inefficient solutions such as centralized deployments \cite{MultiDroneDeployment, CSCA-ADMM} in which an aggregation center coordinates the operations of the UAV-relays; or either joint multi-relay optimization methods \cite{CSCA-ADMM, UAVDynamicCoverage} or model-free formulations consisting of combined state and action spaces \cite{DDQN, MEC-DDPG}. Centralized swarm deployments bring in the need for additional CAPEX and OPEX; and joint multi-UAV constructions lead to prohibitively large solution spaces resulting in unnecessary overhead in policy convergence times, which when scaled to larger swarms result in intractability. Thus, we present an orchestration framework suitable for decentralized UAV-relay swarms by embedding our single UAV-relay policy with multi-agent heuristics and replicating it across the swarm.

A previous version of this research \cite{ICC} focused only on single UAV-relay deployments, assumed an FSPL channel model, and employed PSO for trajectory optimization: the challenges associated with these have been discussed above.

\noindent{\textbf{Novelties}}: In this paper, with rate adaptation to exploit A2G channel stochastics, we first constrain our study to single relay settings, wherein the problem of minimizing the time-average service delay subject to an average UAV power constraint, is formulated as a Semi-Markov Decision Process (SMDP). We derive a multiscale decomposition to this formulation: optimizing the long-term delay-power costs yields outer decisions on radial wait velocities and service positions (via value iteration); consequently, greedily minimizing the instantaneous delay-power costs yields inner actions on angular wait velocities (via exhaustive search) and service trajectories (via competitive swarm optimization). Post single relay policy convergence, with an overlaid command-and-control network, we supplement this control strategy with multi-agent heuristics---namely, spread maximization and consensus-driven conflict resolution---and replicate it across the swarm.

\noindent{\textbf{Extensions}}: Further developments to this research include M/G/$x$ queue management heuristics for link-layer prescient scheduling, a hierarchical variant of CSO to facilitate efficient scalability to higher-dimensional trajectory design, and viability analyses via emulations and real-world flight-tests on NSF AERPAW (OFDM PHY radio + MAVLink vehicle control).

The rest of the paper is structured as follows: Sec. \ref{II} outlines the system model; Sec. \ref{III} elucidates our SMDP formulation for single relay settings; Sec. \ref{IV} details our swarm extensions; Sec. \ref{V} chronicles our numerical evaluations; Sec. \ref{VI} lists our conclusions.

\vspace{-8mm}
\section{System Model}\label{II}
In this section, we model the processes involved in the non-terrestrial augmentation of conventional radio ecosystems, aided in their coverage and service capabilities by UAV-relays.\\
\noindent{\textbf{Deployment Model}}: Consider a generalized deployment in which a swarm of $N_{U}$ rotary-wing Unmanned Aerial Vehicles (UAVs)---each equipped with an on-board transceiver chain---operate as cellular relays to supplement the coverage and service capabilities of a terrestrial Base Station (BS) by relaying data traffic dynamically-generated by Ground Nodes (GNs). The BS is located at the center of the circular cell of radius $a$, at height $H_{B}$, while the UAVs operate at a fixed height of $H_{U}$. The GNs are distributed uniformly at random throughout the cell, with a density of $\lambda_{G}$ [GNs per unit area]. The BS utilizes $k$ orthogonal channels to serve the GNs simultaneously via an Orthogonal Frequency Division Multiple Access (OFDMA) strategy; on the other hand, the UAV-relays are restricted to serve one GN at a time through a decode-and-forward scheme. All channels are assumed to have a bandwidth of $B$. Without any loss of generality, studying uplink transmissions only, these GNs generate random data traffic that is to be transmitted to the BS, either directly or by using one of the UAVs in the swarm as a relay.\\
\noindent{\textbf{Communication Model}}: Each GN generates uplink transmission requests of $L$ bits, according to a Poisson process with rate $\lambda_{R{|}G}$ [requests per GN per unit time]. Coupled with the random deployment of GNs, uplink requests arrive in time according to a Poisson process with rate $\lambda_{R}{=}\lambda_{G}{\cdot}\lambda_{R{|}G}$ [requests per unit time per unit area]. Thus, $\Lambda{\triangleq}\lambda_{R}\pi a^{2}$ [requests per unit time] is the overall request arrival rate over the circular cell. Since a new request is uniformly distributed in the cell area, its angular coordinate $\theta$ is uniform in $[0,2\pi)$ and the probability density function of its radial coordinate is $f_{R}(r){=}\frac{2r}{a^2}\mathbb{I}(r{\leq}a)$, where $\mathbb{I}(\cdot)$ is the indicator function. Allocating the band-edges of the spectrum under use as control channels, a fully-connected mesh network is overlaid on the BS and the UAVs to establish a command-and-control network. Since the packets exchanged among the mesh nodes over the control channel constitute short frames relative to the large data payloads generated by the GNs (and communicated over orthogonal data channels), it is reasonable to neglect the latencies involved in these control operations. When a GN decides to upload its data, it informs the BS---over the control channel---about the need for an uplink transmission of $L$ bits, and includes its physical location in this preliminary request for service. Considering potential delay-power costs for this request, the BS and the UAVs coordinate over the control network to arrive at a consensus on the best scheduling decision: if direct transmission is chosen, the BS assigns a data channel $k{\in}\{1,2,{\dots},N_{B}\}$ to the GN and instructs it to begin transmission; else, if relaying the data payload through UAV $i$ is determined to be the most efficient choice, the UAV instructs the GN to begin transmission over its designated pre-determined data channel $k_{U_{i}}{\in}\{1,2,{\dots},N_{U}\}$. A Decode-and-Forward (D\&F) strategy underlies the communication process encountered in the latter case: while moving along a designed trajectory (a sequence of way-points and velocities), UAV $i$ first receives the entire data payload from the GN over channel $k_{U_{i}}$ (decode) and subsequently transmits it to the BS over the same channel (forward). Crucially, inherent in these scheduling decisions is the A2G channel model underlying the GN$\rightarrow$BS, GN$\rightarrow$UAV, and the UAV$\rightarrow$BS links.\\
\noindent{\textbf{A2G Channel Model}}: For a generic link, we denote the flat-fading channel coefficient as $h{\triangleq}\sqrt{\beta}g$, where $\beta$ captures the large-scale channel variations, and $g$ with $\mathbb{E}\left[|g|^2\right]{=}1$ is the small-scale fading component. We model the large-scale component as $\beta{=}\beta_{\mathrm{LoS}}(d){\triangleq}\beta_{0}d^{-\alpha}$ for line-of-sight (LoS) and $\beta{=}\beta_{\mathrm{NLoS}}(d){\triangleq}\kappa\beta_{0}d^{-\tilde{\alpha}}$ for non-LoS (NLoS) links, where $\beta_{0}$ is the pathloss referenced at a distance of $1$ m, $2{\leq}\alpha{\leq}\tilde{\alpha}$ are the LoS and NLoS path-loss exponents, $\kappa{\in}(0,1]$ captures the additional NLoS attenuation, and $d$ is the Euclidean distance between the transmitter (Tx) and the receiver (Rx) \cite{SCA}. We model the LoS and NLoS probability as a function of the elevation angle $\varphi{\in}(0,90^{o}]$, i.e., 
\begin{align}\label{eq:PLoS}
	P_{\mathrm{LoS}}(\varphi){=}\frac{1}{1{+}z_{1}e^{{-}z_{2}[\varphi{-}z_{1}]}};P_{\mathrm{NLoS}}(\varphi){=}1{-}P_{\mathrm{LoS}}(\varphi),
\end{align}
where $z_{1}$ and $z_{2}$ are environment-specific parameters. The distribution of the small-scale fading component $g$ also depends on the LoS or NLoS link state---specifically, for LoS, we model $g$ as Rician fading with a $\varphi$-dependent $K$-factor, i.e., $K(\varphi){=}k_{1}\exp\left(k_{2}\varphi\right)$, where coefficients $k_{1}$ and $k_{2}$ are determined by the propagation environment \cite{Rician}; for NLoS, we model $g$ as Rayleigh fading (Rician with $K{=}0$) \cite{Rician}. Given $h$, the link capacity is $C(h){=}B{\cdot}\log_{2}\left(1{+}\frac{|h|^{2}P_{T}}{N_{0}B\Gamma} \right)$, where $P_{T}$ is the transmission power, $N_{0}$ is the noise power spectral density at the receiver, $B$ is the channel bandwidth, and $\Gamma$ is the Signal-to-Noise Ratio (SNR) gap between practical modulation-and-coding schemes and theoretical Gaussian signaling \cite{Rician}. We assume that other sources of signal degradation, such as the Doppler effect, are well-compensated at the receiver \cite{Doppler}. Since the large-scale components typically vary slowly relative to the rate of acquisition of Channel State Information (CSI), we assume that the current large-scale parameters $(\beta,K)$ are known at the transmitter's side throughout the communication process, which enables rate control at the transmitter; on the other hand, small-scale fading conditions vary on a much faster timescale, hence cannot be tracked at the transmitter, which may result in outages when the selected rate exceeds the channel capacity $C(h)$. Thus, given $(\beta,K)$ and a transmission rate of $\Upsilon$ [bits per second], we define the outage probability $P_{\mathrm{out}}(\Upsilon,\beta,K) \triangleq \mathbb{P}(C(\sqrt{\beta}g){<}\Upsilon)|\beta,K) = \mathbb{P}\left(|g|^{2}{<}u(\Upsilon,\beta)\right)$, where $u(\Upsilon,\beta){\triangleq}N_{0}B\Gamma(2^{\Upsilon/B}{-}1)/(\beta P_{T})$. Since $2(K{+}1)|g|^{2}$ has a non-central $\chi^2$ distribution with $2$ degrees of freedom and a non-centrality parameter $2K$, we can write the outage probability as
\begin{align}
	P_{\mathrm{out}}(\Upsilon,\beta,K)=1-Q_{1}\left(\sqrt{2K},\sqrt{2(K+1)u(\Upsilon,\beta)}\right),
\end{align}
where $Q_{1}(\cdot,\cdot)$ is the standard Marcum $Q$-function \cite{Rician}. Note that when $K{=}0$ (Rayleigh fading NLoS link), the function specializes to $Q_{1}\left(0,\sqrt{2u(\Upsilon,\beta)}\right){=}\exp(-u(\Upsilon,\beta))$. We assume that the small-scale fading is averaged out across time and space, yielding the expected throughput
\begin{align}
	R(\Upsilon,\beta,K)=\Upsilon\cdot Q_{1}\left(\sqrt{2K},\sqrt{2(K+1)u(\Upsilon,\beta)}\right).
\end{align}
In our model, we permit rate adaptation at the transmitter based on the large-scale parameters $(\beta,K)$, coordinated through the control channel via CSI feedback. The transmission rate $\Upsilon$ is chosen to maximize the expected throughput given $(\beta,K)$, i.e., $\Upsilon^{*}(\beta,K){\triangleq}\argmax_{\Upsilon{\geq}0}R(\Upsilon,\beta,K)$. Let $Z{\triangleq}\sqrt{\frac{2{\beta}P_{T}}{N_{0}B\Gamma}u(\Upsilon,\beta)}$, so $\Upsilon{=}B\log_{2}\left(1{+}\frac{1}{2}Z^{2}\right){\triangleq}f(Z)$,
\begin{align}
    &\Upsilon^{*}(\beta,K){=}f(Z^{*}(\beta,K)),\ Z^{*}(\beta,K){\triangleq}\argmin_{Z{\geq}0}g(Z),\nonumber\\
    &g(Z){\triangleq}-\ln f(Z){-}\ln Q_{1}\left(\sqrt{2K},\sqrt{\frac{(K{+}1)N_{0}B\Gamma}{\beta P_{T}}}Z\right).
\end{align}
Since the function $g(Z)$ is convex, $Z^{*}(\beta,K)$ can be found efficiently using a bisection method. Upon determining the optimal transmission rate $\Upsilon^{*}(\beta,K)$, we define the optimized throughput, as a function of the large-scale conditions, as $R^{*}(\beta,K) \triangleq R(\Upsilon^{*}(\beta,K),\beta,K)$. Assuming that the LoS and NLoS conditions are averaged out in the temporal and spatial dimensions, we compute the average link throughput coupled with rate adaptation as
\begin{align}\label{TBar}
	\bar{R}(d,\varphi)\ \triangleq\ &P_{\mathrm{LoS}}(\varphi)\cdot R^{*}(\beta_{\mathrm{LoS}}(d),K(\varphi))\ +\nonumber\\
	&P_{\mathrm{NLoS}}(\varphi)\cdot R^{*}(\beta_{\mathrm{NLoS}}(d),0),
\end{align}
which is then specialized to the three distinct communication links by expressing the transmission powers, the environment-specific parameters $z_{1}$, $z_{2}$, $k_{1}$, and $k_{2}$, the large-scale parameters $(\beta,K)$, and the LoS or NLoS probabilities \eqref{eq:PLoS} based on the spatial configuration, i.e., Tx-Rx distance and elevation angle. Specifically, for the GN$\rightarrow$BS link, we let $\bar{R}_{GB}(r)$ be the throughput with the GN in position $(r,\theta)$, computed by setting the GN-BS distance as $d{=}\sqrt{H_{B}^{2}{+}r^{2}}$ and the elevation angle as $\varphi{=}\sin^{-1}\left(H_{B}/d\right)$ in \eqref{TBar}. Similarly, for the GN$\rightarrow$UAV link, we let $\bar{R}_{GU}(r_{GU})$ be the throughput when the GN-UAV distance (projected onto the $x{-}y$ plane) is $r_{GU}$, computed by setting the GN-UAV Euclidean distance as $d{=}\sqrt{r_{GU}^{2}{+}H_{U}^{2}}$ and the elevation angle as $\varphi{=}\sin^{-1}\left(H_{U}/d\right)$ in \eqref{TBar}. Finally, for the UAV$\rightarrow$BS link, we let $\bar{R}_{UB}(r_{UB})$ be the throughput when the $x{-}y$ projected UAV-BS distance is $r_{UB}$, computed by setting the GN-UAV Euclidean distance as $d{=}\sqrt{r_{UB}^{2}{+}(H_{U}{-}H_{B})^{2}}$ and $\varphi{=}\sin^{-1}\left(\frac{(H_{U}{-}H_{B})}{d}\right)$ in \eqref{TBar}.\\

\vspace{-6mm}
\section{The SMDP Formulation}\label{III}
\begin{figure} [t]
    \centering
    \includegraphics[width=1.0\linewidth]{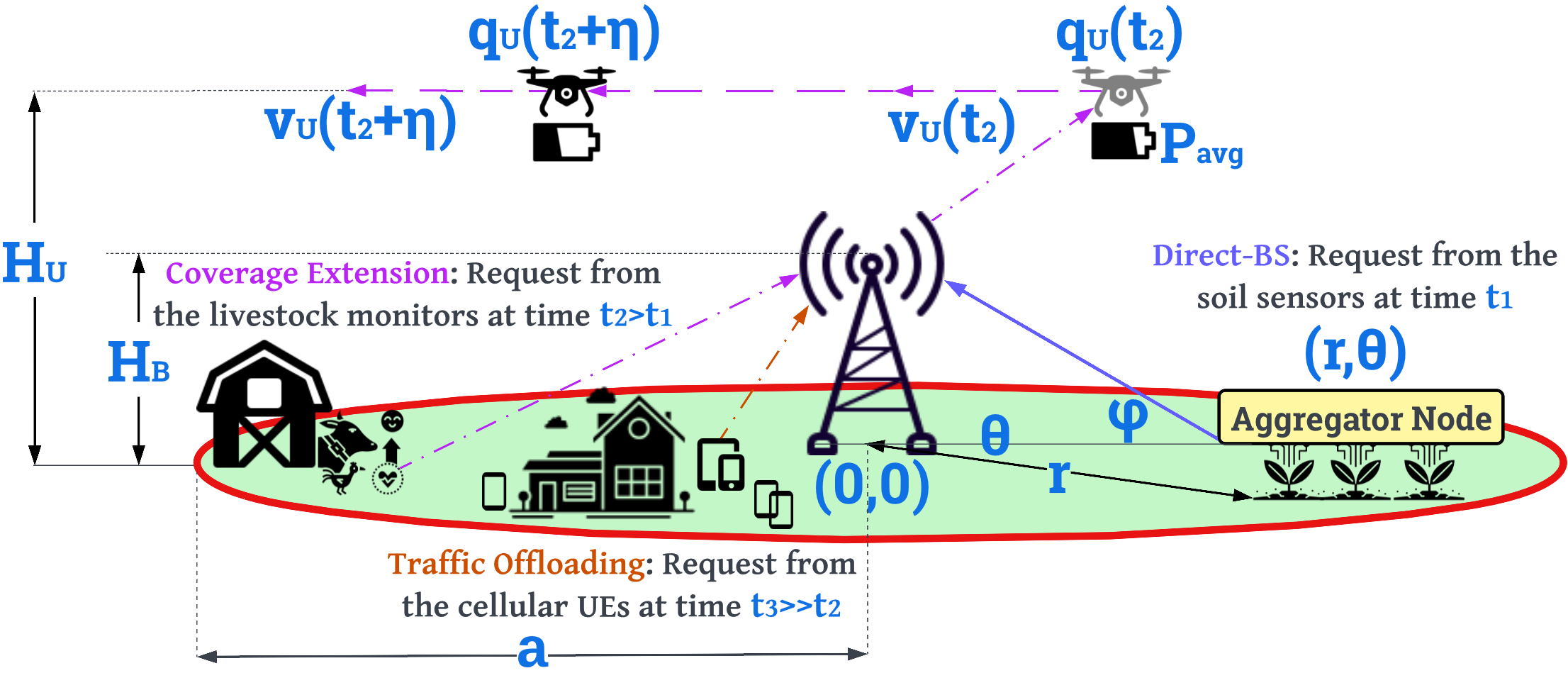}
    \vspace{-6mm}
    \caption{Single UAV-relay specialization of our generalized deployment setting.}
    \vspace{-6mm}
    \label{F4}
\end{figure}
In this section, we specialize the generalized deployment, communication, and channel models detailed in Sec. \ref{II} to single UAV-relay settings. Accordingly, we describe the mathematical constructions involved in the design of our solution framework to minimize the time-average service delay experienced by the GNs in the cell, subject to an average UAV mobility power constraint via a Semi-Markov Decision Process (SMDP) formulation. The effective traffic rate experienced by a single UAV is $\frac{\Lambda}{N_U}$ [requests per unit time], which is assumed in this section in place of the overall rate $\Lambda$. Let $\mathbf{q}_{U}(t){=}(r_{U}(t),\theta_{U}(t))$ be the polar coordinate of the UAV at time $t$, projected onto the $x{-}y$ plane, where $r_{U}(t){\in}[0,a]$ and $\theta_{U}(t){\in}[0,2\pi)$ denote the UAV's radius and angle with respect to the BS (cell center). This setup is depicted in Fig. \ref{F4}.

We note that the operations of the UAV-relay can be split into the following phases. In the waiting phase, no GN requests are being served by the UAV, which thus moves according to a waiting policy, until a new request is received. When a new GN request is received, say from position $(r,\theta)$, the system transitions to the request scheduling phase, where the system decides if the GN should transmit its data payload directly to the BS, or relay it through the UAV. If direct transmission is selected, the system immediately re-enters the waiting phase, as the UAV remains free to serve other requests; else, the system enters the UAV-relay phase, in which the GN relays its data payload through the UAV using the D\&F protocol; upon the completion of this relay service, the system re-enters the waiting phase. Note that the BS can accommodate simultaneous transmissions (see Sec. \ref{II}): new requests received during the UAV-relay phase are directly served by the BS. Under a given policy $\mu$, averaged out over $M_{t}$ decision intervals, we denote the expected average service delay for scheduled requests as $\bar{W}_{\mu}^{(s)}$, the expected average UAV energy expenditure as $\bar{E}_{\mu}$, and the expected average operational duration as $\bar{T}_{\mu}$. With this setup, the optimization problem given by ${\mathrm{min}}_{\mu}\bar{W}_{\mu}^{(s)}$ s.t. $\bar{E}_{\mu}{-}P_{\mathrm{avg}}\bar{T}_{\mu}{\leq}0$, will be the subsequent focus of our analyses. To solve it, consider
\begin{align}\label{eq:W2Lagr}
    g(\nu)&{=}\underset{\mu}{\mathrm{min}}\ \bar{W}_{\mu}^{(s)} + \nu (\bar{E}_{\mu}-P_{\mathrm{avg}}\bar{T}_{\mu}),
\end{align}
where $\nu$ is the dual variable, optimized by solving $\max_{\nu{\geq}0}g(\nu)$. We now demonstrate that for a given $\nu{\geq}0$, \eqref{eq:W2Lagr} can be cast as an SMDP and solved via dynamic programming.\\
\noindent{\textbf{States}}: The state is defined by the UAV position $\mathbf{q}_{U}$, taking value from the set $\mathcal{Q}_{\mathrm{UAV}}\triangleq[0,a]\times[0,2\pi)$ (polar coordinates), and the position of an uplink transmission request $\mathbf{q}_G$, taking values from $\mathcal{Q}_{\mathrm{GN}}{\triangleq}[0,a]{\times}[0,2\pi)$ (polar coordinates). The state space is then $\mathcal{S}{=}\mathcal{S}_{\mathrm{wait}}{\cup}$ $\mathcal{S}_{\mathrm{comm}}$, where $\mathcal{S}_{\mathrm{wait}}{=}\mathcal{Q}_{\mathrm{UAV}}$ is the set of waiting states and $\mathcal{S}_{\mathrm{comm}}{=}\mathcal{Q}_{\mathrm{UAV}}{\times}\mathcal{Q}_{\mathrm{GN}}$ is the set of communication states. Crucial to the definition of the SMDP is how the system is sampled in time to define Markovian dynamics in the evolution of the sampled states. Accordingly, next, we define the actions available in each state $\mathbf{s}{\in}\mathcal{S}$ and the transition probabilities, along with the duration $T(\mathbf{s};\mathbf{a})$, the UAV energy usage $E(\mathbf{s};\mathbf{a})$, and the communication delay $\Delta(\mathbf{s};\mathbf{a})$ metrics accrued in state $\mathbf{s}$ under action $\mathbf{a}$.\\
\noindent{\textbf{Waiting actions and transitions}}: If the UAV is in the waiting state $\mathbf{s}_{n}{=}\mathbf{q}_{U}{\in}\mathcal{S}_{\mathrm{wait}}$ at time $t$, then the actions available are to move the UAV with radial (referred outward) and angular (referred counter-clockwise) velocity components $(v_{r},\theta_{c})$, over an arbitrarily small duration $\Delta_{0}{\ll}1/\Lambda$. Under a maximum velocity constraint $V_{\mathrm{max}}$, the action space is then $\mathcal{A}_{\mathrm{wait}}(r_{U}){\triangleq}\Big\{(v_{r},\theta_{c}){\in}\mathbb{R}^{2}\Big|\sqrt{v_{r}^{2}{+}r_{U}^{2}{\cdot}\theta_{c}^{2}}{\leq}V_{\mathrm{max}}\Big\}$, where $v_{U}{=}\sqrt{v_{r}^{2}{+}r_{U}^{2}\theta_{c}^{2}}$ is the velocity expressed with respect to polar coordinates. Upon choosing action $\mathbf{a}{=}(v_{r},\theta_{c}){\in}\mathcal{A}_{\mathrm{wait}}(r_{U})$, the communication delay is $\Delta(\mathbf{s};\mathbf{a}){=}0$, since there is no ongoing communication; the duration of a waiting state visit is $T(\mathbf{s};\mathbf{a}){=}\Delta_{0}$, during which the UAV uses an amount of energy $E(\mathbf{s};\mathbf{a}){=}\Delta_{0}P_{\mathrm{mob}} \left(v_{U}\right)$ to move at velocity $v_{U}$. The new state is then sampled at time $t{+}\Delta_{0}$, with the UAV moved to the new position $\mathbf{q}_{U}(t{+}\Delta_{0}){\approx}(r_{U},\theta_{U}){+}(v_{r},\theta_{c})\Delta_{0}$. With probability $e^{-\Lambda \Delta_{0}}$, no new request is received in the time interval $[t,t{+}\Delta_{0}]$, so that the new state is a waiting state.  Otherwise, a new request is received from a GN in position $(r,\theta)$, so that the new state is a communication state. Thus, the transition probability from the waiting state $\mathbf{s}_{n}{=}\mathbf{q}_{U}$ under action $\mathbf{a}_{n}{=}(v_{r},\theta_{c}){\in}\mathcal{A}_{\mathrm{wait}}(r_{U})$ is 
\begin{align}\label{eq:R0ContTrProb}
    &\mathbb{P}(\mathbf{s}_{n+1}{=}\mathbf{q}_{U}{+}\mathbf a_{n}\Delta_{0}{|}\mathbf{s}_{n},\mathbf{a}_{n}){=}e^{-\Lambda \Delta_{0}}\text{ and }\\
    &\mathbb{P}(\mathbf{s}_{n{+}1}{=}(\mathbf{q}_{U}{+}\mathbf{a}_{n}\Delta_{0},\mathbf{q}_{G})|\mathbf{s}_{n},\mathbf{a}_{n}){=}\frac{A(\mathcal{F})}{\pi a^{2}}{\cdot}(1{-}e^{-\Lambda\Delta_{0}}),\nonumber
\end{align}
where $\mathbf{q}_{G}{\in}\mathcal{F}$ and $A(\mathcal{F})$ is the area of region $\mathcal{F}$, ${\forall}\mathcal{F}{\subseteq}\mathcal{Q}_{\mathrm{GN}}$, since requests are uniformly distributed in the cell.\\
\noindent{\textbf{Communication actions and transitions}}:
Upon reaching a communication state $\mathbf{s}_{n}{=}(\mathbf{q}_{U},\mathbf{q}_{G}){\in}\mathcal{S}_{\mathrm{comm}}$ at time $t$, the system must serve a GN request at position $\mathbf{q}_{G}{=}(r,\theta)$. The system first determines the best scheduling decision $\xi{\in}\{0,1\}$. If $\xi{=}0$, the GN transmits directly to the BS, and the next state immediately after this decision is the waiting state $\mathbf{s}_{n{+}1}{=}\mathbf{q}_{U}{\in}\mathcal{S}_{\mathrm{wait}}$ with probability $1$. In this case, the cost metrics under action $\mathbf{a}_{n}{=}\mathbf{a}{=}(0,[\ ])$ are computed as $\Delta(\mathbf{s}_{n};\mathbf{a}){=}\frac{L}{\bar R_{GB}(r)}$, $E(\mathbf{s}_{n};\mathbf{a}){=}0$, $T(\mathbf{s}_{n};\mathbf{a}){=}0$, since direct transmissions occur at throughput $\bar{R}_{GB}(r)$ and the system moves immediately to the waiting state $\mathbf{q}_{U}{\in}\mathcal{S}_{\mathrm{wait}}$ resulting in the action duration and energy expenditure being $0$. On the other hand, if $\xi{=}1$, the UAV uses the D\&F protocol described next, while following a trajectory starting from its current position $\mathbf{q}_{U}$ and ending in position $\mathbf{q}_{U}'$. We denote this action as $\mathbf{a}_{n}{=}\mathbf{\tilde{a}}{=}(1,\mathbf{q}_{U}{\rightarrow}{\mathbf{q}}_{U}')$. In the first phase (of duration $t_{p}$) of the D\&F protocol, the GN transmits its payload to the UAV; in the second phase (of duration $\Delta{-}t_{p}$), the UAV relays the data payload to the BS. Assuming a move-and-transmit implementation \cite{SCA}, the trajectory ($\mathbf{q}_{U}{\rightarrow}{\mathbf{q}}_{U}'$) and the time periods ($T_{1}{\triangleq}[0{,}t_{p}]$ and $T_{2}{\triangleq}(t_{p}{,}\Delta]$) must satisfy
\begin{align}
	\int_{T_{1}}\bar{R}_{GU}(r_{GU}(t{+}\eta))\mathrm{d}{\eta},\int_{T_{2}}\bar{R}_{UB}(r_{UB}(t{+}\eta))\mathrm{d}{\eta}{\geq}L\label{eq:PLConst1}\tag{C.1},
\end{align}
i.e., the entire payload of $L$ bits is first transmitted to the UAV with rate $\bar{R}_{GU}(r_{GU}(t{+}\eta))$, where $r_{GU}(t{+}{\eta})$ is the GN-UAV distance (projected onto the $x{-}y$ plane) at time $t{+}\eta$; then, the UAV transmits the payload to the BS with rate $\bar{R}_{UB}(r_{UB}(t{+}\eta))$, where $r_{UB}(t{+}\eta)$ is the radial position of the UAV at time $t{+}\eta$, so that the total communication delay is $\Delta$.  In this case, the cost metrics under action $\mathbf{\tilde{a}}$ are $\Delta(\mathbf{s}_{n};\mathbf{\tilde{a}}){=}\Delta$, $E(\mathbf{s}_{n};\mathbf{\tilde{a}}){=}\int_0^\Delta P_{\mathrm{mob}}\left(v_{U}(t{+}\eta)\right)\mathrm{d}\eta$, and $T(\mathbf{s}_{n};\mathbf{\tilde{a}}){=}\Delta$. Upon completing the D\&F protocol, the UAV enters the waiting phase again, so that $\mathbf{s}_{n{+}1}{=}\mathbf{q}_{U}'$ becomes the new SMDP state, sampled at time $t{+}\Delta$. Let $\mathcal{Q}_{\mathbf{q}_{G}}\big(\mathbf{q}_{U}{\rightarrow}{\mathbf{q}}_{U}'\big)$ be the set of feasible UAV trajectories starting in $\mathbf{q}_{U}$, terminating in $\mathbf{q}_{U}'$, to serve a GN located at $\mathbf{q}_{G}$ using D\&F, i.e., 
\begin{align*}
	&\mathcal{Q}_{\mathbf{q}_{G}}\big({\mathbf{q}}_{U}\rightarrow{\mathbf{q}}_{U'}\big){=}\Big\{\mathbf{p}_{U}:[0,\Delta]{\mapsto}[0,a]\times[0,2\pi)\text{ s.t. \ref{eq:PLConst1},}\\
	&\mathbf{p}_{U}(0){=}{\mathbf{q}}_{U},\mathbf{p}_{U}(\Delta){=}{\mathbf {q}}_{U}',{\exists}\Delta{\geq}0,{\exists}0{\leq}t_p{\leq}\Delta,\label{eq:IFConst1}\tag{C.2}\\
	&v_{U}(\eta){\leq}V_{\mathrm{max}},{\forall}\eta{\in}[0,\Delta]\label{eq:SpeedConst1}\tag{C.3}\Big\},
\end{align*}
where \ref{eq:IFConst1} reflects the trajectory constraints and \ref{eq:SpeedConst1} reflects the maximum velocity constraint. Then, the action space in state $(\mathbf{q}_{U},\mathbf{q}_{G}){\in}\mathcal{S}_{\mathrm{comm}}$ when $\xi{=}1$ is the set $\mathcal{Q}_{\mathbf{q}_{G}}(\mathbf{q}_{U}){\triangleq}\cup_{\mathbf{q}_{U}'{\in}\mathcal{Q}_{\mathrm{UAV}}}\mathcal{Q}_{\mathbf{q}_{G}}\big(\mathbf{q}_{U}{\rightarrow}\mathbf{q}_{U}'\big)$ of feasible trajectories starting in $\mathbf{q}_{U}$ that serve the GN at $\mathbf{q}_{G}$ via the D\&F protocol. The overall communication action space is then $\mathcal{A}_{\mathrm{comm}}(\mathbf{q}_{U},\mathbf{q}_{G}){\triangleq}\{0,[\ ]\}{\cup}\{\{1\}{\times}\mathcal{Q}_{r,\theta}(r_{U},\theta_{U})\}$. Here the set $\{0, [\ ]\}$ is associated with $\xi{=}0$ (no trajectory design space); while the set $\{\{1\}{\times}\mathcal{Q}_{r,\theta}(r_{U},\theta_{U})\}$ is associated with $\xi{=}1$, whose trajectory design space is $\mathcal{Q}_{r,\theta}(r_{U},\theta_{U})$.\\
\noindent{\textbf{Policy $\mu$}}: For waiting states $\mathbf{q}_{U}{\in}\mathcal{S}_{\mathrm{wait}}$, the policy selects a velocity $(v_{r},\theta_{c})$ from the waiting action space, i.e., $\mu(\mathbf{q}_{U}){\in}\mathcal{A}_{\mathrm{wait}}(r_{U})$. Likewise, for communication states $(\mathbf{q}_{U},\mathbf{q}_{G}){\in}\mathcal{S}_{\mathrm{comm}}$, the policy selects the scheduling decision $\xi{\in}\{0,1\}$ and if $\xi{=}1$, the trajectory followed in the D\&F protocol, i.e., $\mu(\mathbf{q}_{U},\mathbf{q}_{G}){\in}\mathcal{A}_{\mathrm{comm}}(\mathbf{q}_{U},\mathbf{q}_{G})$. With a stationary policy $\mu$ defined, the Lagrangian metric $L_{\mu}^{(\nu)}{\triangleq}\bar{W}_{\mu}^{(s)}{+}\nu(\bar{E}_{\mu}{-}P_{\mathrm{avg}}\bar{T}_{\mu})$ in \eqref{eq:W2Lagr} is reformulated using Little's Law as
\begin{align}\label{eq:CostMetric}
    L_\mu^{(\nu)}
    &= \lim_{N \rightarrow \infty} \mathbb{E}_\mu \Bigg[ \frac{\frac{1}{N}\sum_{n=0}^{N-1}  \ell_\nu(\mathbf{s}_n; \mu(\mathbf{s}_n)) }{\frac{1}{N}\sum_{n = 0}^{N-1} \mathbb I(\mathbf{s}_n \in \mathcal{S}_{\mathrm{comm}})}  \Bigg]\nonumber\\
    &= \frac{1}{\pi_{\mathrm{comm}}}\int_{\mathcal{S}} \Pi_{\mu}(\mathbf{s})\ell_\nu(\mathbf{s}; \mu(\mathbf{s}))\mathrm{d}\mathbf{s},
\end{align}
where $\Pi_{\mu}(\mathbf{s})$ is the steady-state probability density function of the SMDP being in a state $\mathbf{s}$ under policy $\mu$, $\pi_{\mathrm{comm}}{=}\int_{\mathcal{S}_{\mathrm{comm}}}\!\!\!\!\!\Pi_{\mu}(\mathbf{s})\mathrm{d}\mathbf{s}$ is the steady-state probability that the UAV is in a communication state of the SMDP, and $\ell_{\nu}(\mathbf{s};\mathbf{a}){\triangleq}\Delta(\mathbf{s};\mathbf{a}){+}\nu\big(E(\mathbf{s};\mathbf{a}){-}P_{\mathrm{avg}}T(\mathbf{s};\mathbf{a})\big)$ is the overall Lagrangian metric in state $\mathbf{s}$ under action $\mathbf{a}$. In \eqref{eq:CostMetric}, note that $\sum_{n=0}^{N{-}1}\ell_{\nu}(\mathbf{s}_{n};\mu(\mathbf{s}_{n}))$ is the total Lagrangian cost accrued during the first $N$ SMDP stages, and $\sum_{n{=}0}^{N{-}1}\mathbb{I}(\mathbf{s}_{n}{\in}\mathcal{S}_{\mathrm{comm}})$ is the number of communication states encountered in the SMDP: since a new decision interval is initiated after a communication state, this in turn equals the number of decision intervals. Therefore, after taking the expectation and the limit $N{\to}\infty$, $L_{\mu}^{(\nu)}$ represents the expected Lagrangian cost per decision interval, as expressed in \eqref{eq:W2Lagr}. The subsequent right hand expression in \eqref{eq:CostMetric} follows by noticing that the SMDP achieves a steady-state behavior when $N\to\infty$. We now specialize the Lagrangian metric $\ell_{\nu}(s;\mathbf{a})$. Specifically, for waiting states,
\begin{align}\label{eq:EllWait}
    \ell_{\nu}(r_{U},\theta_{U};v_{r},\theta_{c}){=}\nu\Big(P_{\mathrm{mob}}\Big(\sqrt{v_{r}^{2}{+}r_{U}^{2}\theta_{c}^{2}}\Big){-}P_{\mathrm{avg}}\Big)\Delta_{0};
\end{align}
for communication states under $\xi{=}0$, $\ell_{\nu}(r_U,\theta_U,r,\theta;0,[\ ])=\frac{L}{\bar{R}_{GB}(r)}$; and for communication states under $\xi{=}1$ with trajectory 
$\mathbf{p}_{U}$ of duration $\Delta$, we obtain
\begin{align*}
    \ell_{\nu}(r_{U}{,}\theta_{U}{,}r{,}\theta{;}1{,}\mathbf{p}_{U}){=}(1{-}\nu P_{\mathrm{avg}})\Delta{+}\nu\int_{0}^{\Delta}P_{\mathrm{mob}}\left(v_{U}(\eta)\right)\mathrm{d}\eta.
\end{align*}
The minimization problem of \eqref{eq:W2Lagr} can then be expressed as the average cost-per-stage problem
\begin{align}\label{eq:TotalGMin}
	g(\nu) = \frac{1}{\pi_{\mathrm{comm}}}\underset{\mu}{\mathrm{min}} \; \int_{\mathcal{S}} \Pi_{\mu}(\mathbf{s}) 
	\ell_\nu(\mathbf{s}; \mu(\mathbf{s}))\mathrm d \mathbf{s},
\end{align}
solvable via standard dynamic programming approaches, after discretization of the state and action spaces, followed by the dual maximization $\mathrm{max}_{\nu{\geq}0}g(\nu)$. Since GN transmission requests are uniformly distributed in the circular cell with the BS in the center, the UAV radius information is a sufficient statistic in decision making for a waiting state $(r_{U},\theta_{U})$, which can be thus expressed as $\mathbf{s}{=}r_{U}{\in}\mathcal{S}_{\mathrm{wait}}$; likewise, for a communication state $(r_{U},\theta_{U},r,\theta)$, only the UAV radius, GN request radius, and the angle $\psi{\in}[0,2\pi)$ between them suffice to characterize the state---thus, communication states can be compactly represented as $\mathbf{s}{=}(r_{U},r,\psi){\in}\mathcal{S}_{\mathrm{comm}}$. A consequence of these sufficient statistics for decision-making is that the policy affects the SMDP state transitions (hence, steady-state behavior) only through the UAV radial velocity $v_{r}$ in the waiting states and the UAV trajectory's target end radius position $\hat{r}_{U}$ in the communication states. On the other hand, the angular velocity $\theta_{c}$ in the waiting states and the UAV trajectory's target end angular coordinate $\hat{\theta}_{U}$ in the communication states do not influence state dynamics, but only the instantaneous Lagrangian metric $\ell_{\nu}$. With this observation, let $O(\mathbf{s}){\triangleq}v_{r}{\in}[-V_{\mathrm{max}},V_{\mathrm{max}}]$ define the radial velocity policy of the waiting states $\mathbf{s}{\in}\mathcal{S}_{\mathrm{wait}}$, specifying the radial velocity component of a waiting action $\mathbf{a}{=}(v_{r},\theta_{c}) \in \mathcal{A}_{\mathrm{wait}}(\mathbf{s})$; let $U(\mathbf{s}){\triangleq}\hat{r}_{U}{\in}[0,a]$ define the next radius position policy of the communication states $\mathbf{s}{\in}\mathcal{S}_{\mathrm{comm}}$, specifying the end radius position of a scheduling and communication action $\mathbf{a}{\in}\mathcal{A}_{\mathrm{comm}}(\mathbf{s})$. Under this decomposition, $O$ and $U$ constitute the outer decisions made by the SMDP and are the only actions to affect the steady-state distribution, denoted as $\Pi_{O,U}$ under the outer policy $(O,U)$. Thus, the optimization problem \eqref{eq:TotalGMin} can be restated as
\begin{align}\label{eq:PolDecomp}
	g(\nu) = \frac{1}{\pi_{\mathrm{comm}}} \underset{O,U}{\mathrm{min}} \Bigr[ &\int_{\mathcal{S}_{\mathrm{wait}}} \Pi_{O,U}(\mathbf{s}) \ell_{\nu}^{*}(\mathbf{s}; O(\mathbf{s}))\mathrm{d}\mathbf{s} +\nonumber\\&\int_{\mathcal{S}_{\mathrm{comm}}} \Pi_{O,U}(\mathbf{s}) \ell_{\nu}^{*}(\mathbf{s}; U(\mathbf{s})) \mathrm{d}\mathbf{s} \Bigr],
\end{align}
where $\ell_{\nu}^{*}$ is the Lagrangian metric optimized with respect to the inner action components not specified by $O$ and $U$. In particular, for waiting states $\mathbf{s}{=}r_{U}$ and radial velocity $O(\mathbf{s}){=}v_{r}$, the inner optimization is performed with respect to the UAV angular velocity $\theta_{c}$, i.e.,
\begin{align}\label{eq:MinLWP}
	\ell_{\nu}^{*}(\mathbf{s}; v_r) = &\underset{\theta_c}{\mathrm{min}}\; \nu \left( P_{\mathrm{mob}}\left(\sqrt{v_{r}^2 + r_U^2\theta_c^2}\right) - P_{\mathrm{avg}} \right)\Delta_0\nonumber\\&\mathrm{s.t.}\;\; \sqrt{v_{r}^{2} + r_U^2\theta_c^2} \leq V_{\mathrm{max}}.
\end{align}
Since $\nu{\geq}0$, $\Delta_{0}{>}0$, and $P_{\mathrm{avg}}$ are constant, the optimizer $\theta_{c}^{*}$ is the angular velocity minimizing the UAV power consumption for a given UAV radial velocity $v_{r}$ and radius $r_{U}$, solvable through exhaustive search. For communication states $\mathbf{s}{=}(r_{U},r,\psi)$, $\ell_{\nu}^{*}(\mathbf{s};\hat{r}_{U})$ is determined by optimizing over $\xi{\in}\{0,1\}$ and if $\xi{=}1$, the trajectory $\mathbf{p}_{U}$ followed by the UAV, terminating in $\hat{r}_{U}$. Let $\ell_{\nu}^{*}(\mathbf{s};\hat{r}_{U},\xi)$ denote the optimized metric as a function of $\xi{\in}\{0,1\}$. For $\xi{=}1$ (D\&F protocol),
\begin{align}\label{eq:EllMin}
    \ell_{\nu}^{*}(\mathbf{s};\hat{r}_{U},1){=}&\underset{\Delta,\mathbf{p}_{U},t_{p}}{\mathrm{min}}(1{-}\nu P_{\mathrm{avg}})\Delta + \\&\nu\int_{0}^{\Delta}P_{\mathrm{mob}}\Bigg(\sqrt{r_{U}^{'}(\eta)^{2}{+}r_{U}^{2}(\eta)\theta_{U}^{'}(\eta)^{2}}\Bigg)\mathrm{d}\eta\text{ s.t. }\nonumber
    \\&\text{\ref{eq:PLConst1}},\text{\ref{eq:SpeedConst1}}\nonumber,\mathbf{p}_{U}(0){=}(r_{U},0),\Vert\mathbf{p}_{U}(\Delta)\Vert_{2}{=}\hat{r}_{U}\tag{$\hat{\text{C}}$.2}\label{eq:StEnd}
\end{align}
where \ref{eq:StEnd} enforces the trajectory constraints. For $\xi{=}0$, $\hat{r}_{U}{=}r_{U}$ and $\ell_{\nu}^{*}(\mathbf{s};r_{U},0){=}\frac{L}{\bar{R}_{GB}(r)}$. Hence $\ell_{\nu}^{*}(\mathbf{s};r_{U})$ is obtained by further minimizing over $\xi{\in}\{0,1\}$, yielding
\begin{align*}\label{ellnushatru}
	\ell_{\nu}^{*}(\mathbf{s};\hat{r}_{U}){=}\underset{\xi{\in}\{0,1\}}{\min}\ell_{\nu}^{*}(\mathbf{s};r_{U},\xi)\mathbb{I}(\hat{r}_{U}{=}r_{U}){+}\ell_{\nu}^{*}(\mathbf{s};\hat{r}_{U},1)\mathbb{I}(\hat{r}_{U}{\neq}{r}_{U}).
\end{align*}
\begin{figure} [t]
    \centering
    \includegraphics[width=1.0\linewidth]{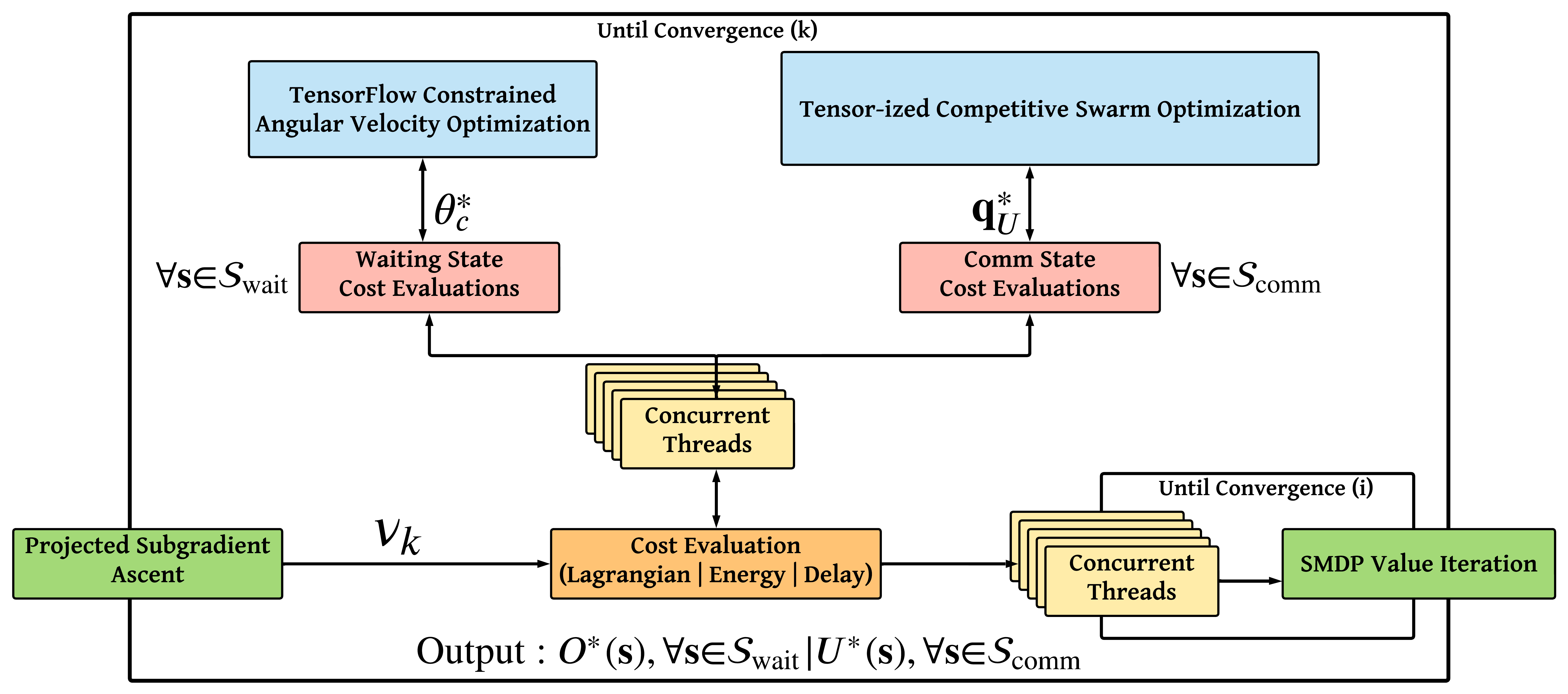}
    \vspace{-5mm}
    \caption{Algorithmic flow inherent in our single UAV policy optimization.}
    \vspace{-7mm}
    \label{F5}
\end{figure}
\noindent Thus, if the outer decision selects $U(\mathbf{s}){=}r_{U}$, the inner scheduling decision $\xi{\in}\{0,1\}$ is obtained by greedily minimizing a cost metric that trades off communication delay and energy consumption, i.e., direct transmission to the BS occurs if $\ell_{\nu}^{*}(\mathbf{s}; r_{U},0){<}\ell_{\nu}^{*}(\mathbf{s};r_{U},1)$. Otherwise, the UAV handles the GN request using the D\&F protocol, and the inner decision on UAV trajectory---designed via Competitive Swarm Optimization (CSO) \cite{CSO}---greedily minimizes the instantaneous delay-power trade-off, terminating at the target radius of the outer decision, $U(\mathbf{s}){=}\hat{r}_{U}$. Discretizing the trajectory between the UAV's initial and final service positions, and using a modified version of \eqref{eq:EllMin} (incorporating time and energy penalties for data transmission failures during D\&F) as the cost function, CSO invokes pair-wise cost comparisons among a randomly initialized set of way-points and velocities: specifically, in each iteration, the winner particles (way-points and velocities with lower cost function values) advance to the next iteration while the loser particle values are updated in relation to these winners; this update process continues until a maximum number of cost evaluations are performed. We design the outer policy and compute the average cost-per-stage metric $g(\nu)$, along with the average energy- and time-per-stage metrics for a given $\nu$, by solving problem \eqref{eq:PolDecomp} via value iteration; also, the dual maximization problem $\mathrm{max}_{\nu{\geq}0}g(\nu)$ is solved via projected sub-gradient ascent \cite{SubgradientMethods}. Fig. \ref{F5} illustrates the sequence of operations involved in solving for the optimal policy.

\vspace{-6mm}
\section{Multi-Agent Extensions}\label{IV}
\begin{figure} [t]
     \centering
     \includegraphics[width=1.0\linewidth]{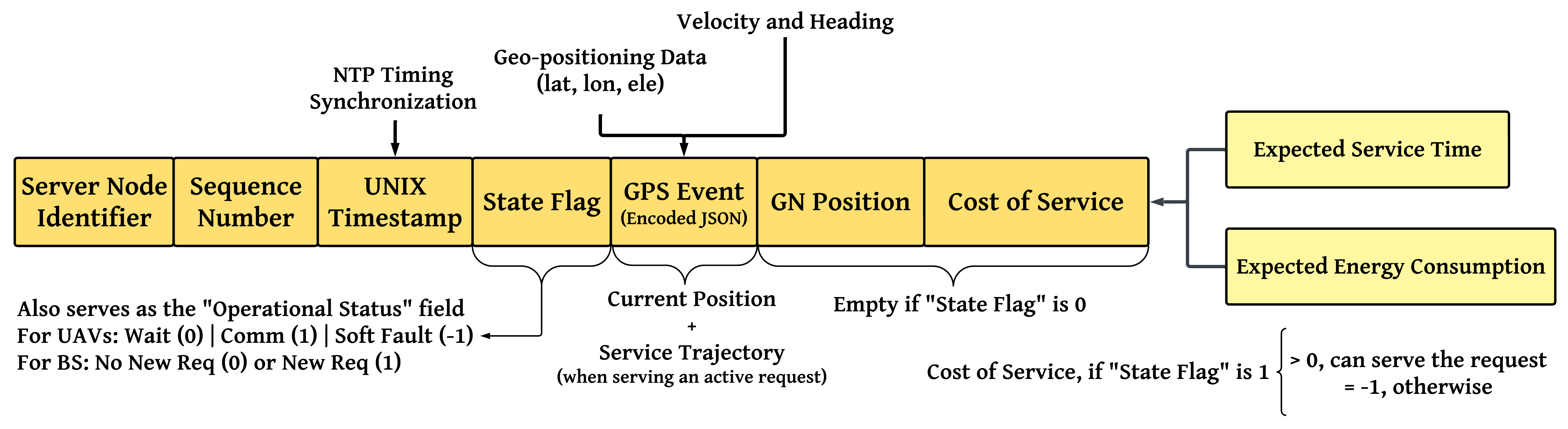}
     \vspace{-5mm}
     \caption{Structure of the control frames exchanged among the BS and UAVs.}
     \vspace{-7mm}
     \label{F6}
\end{figure}

In this section, we remove the specializations considered for our single relay construction in Sec. \ref{III}, and extend the SMDP-based control strategy to our generalized hybrid wireless network model of $N_{U}$ UAV-relays in the swarm. To this end, the single relay optimal policy is embedded with supplementary heuristics and replicated across the swarm.\\
\noindent{\textbf{Command-and-Control}}: Truly decentralized and coordinated operations of the UAV-relays in the swarm necessitates the need for a control network over which the server nodes can collaborate to ensure collision-free movements among the UAVs, facilitate consensus-driven decision-making on the best server to handle a GN uplink transmission request, and guarantee resilient fault-tolerant operations by setting-up fallback mechanisms to handle UAV failures. We designate the band-edges of the allocated spectrum as control channels, over which the server nodes in the cell exchange short collaboration messages: the structure of a control frame in our system is shown in Fig. \ref{F6}. A fully-connected distributed mesh topology (employing these designated control channels) overlaid over the BS and the UAV swarm constitutes the design of our command-and-control network. Since each UAV-relay in the swarm possesses the same optimal waiting and communication state policies, we embed spread maximization (in the waiting states) and conflict resolution (in the communication states) to cooperatively handle their operations.\\
\noindent{\textbf{Spread Maximization}}: To efficiently position and prime the idle UAVs for a potential new GN request, in the waiting states, a UAV-relay in the swarm, in addition to executing the optimal action, determines the direction of its angular motion (clockwise or counter-clockwise) based on our spread maximization heuristic---wherein each UAV-relay in the waiting state executes either positive (counter-clockwise) or negative (clockwise) angular movements in order to maximize the minimum distance among them. These coordinated movements among the UAVs is made possible through periodic exchanges of control frames over the command-and-control network. Studying this frame structure in Fig. \ref{F6}, we note that for waiting states, the state flag is set to $0$, and the GN position and the cost-of-service fields are empty; UAVs in the waiting state extract the positional information of their peers from the GPS event field and apply a maximize-the-minimum-distance heuristic over other idle UAVs in the cell. This methodology ensures that a suitable spread is maintained among the UAVs in the waiting states, in order to facilitate faster response times when a new uplink request is generated.
\begin{figure*} [t]
     \begin{subfigure}{0.5\linewidth}
         \centering
         \includegraphics[width=1.0\linewidth]{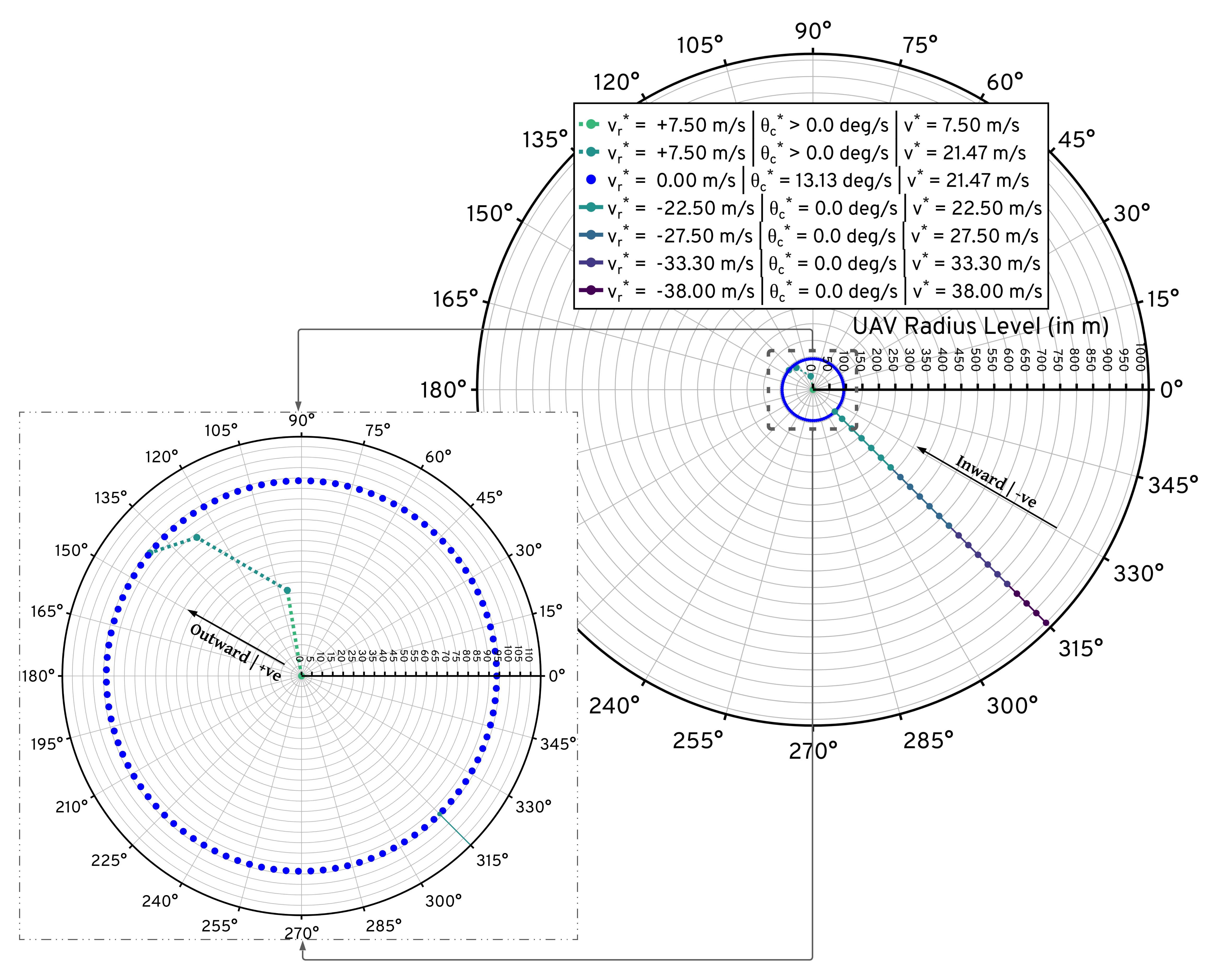}
         \vspace{-4mm}
         \label{F7}
     \end{subfigure}
     \begin{subfigure}{0.5\linewidth}
         \centering
         \includegraphics[width=1.0\linewidth]{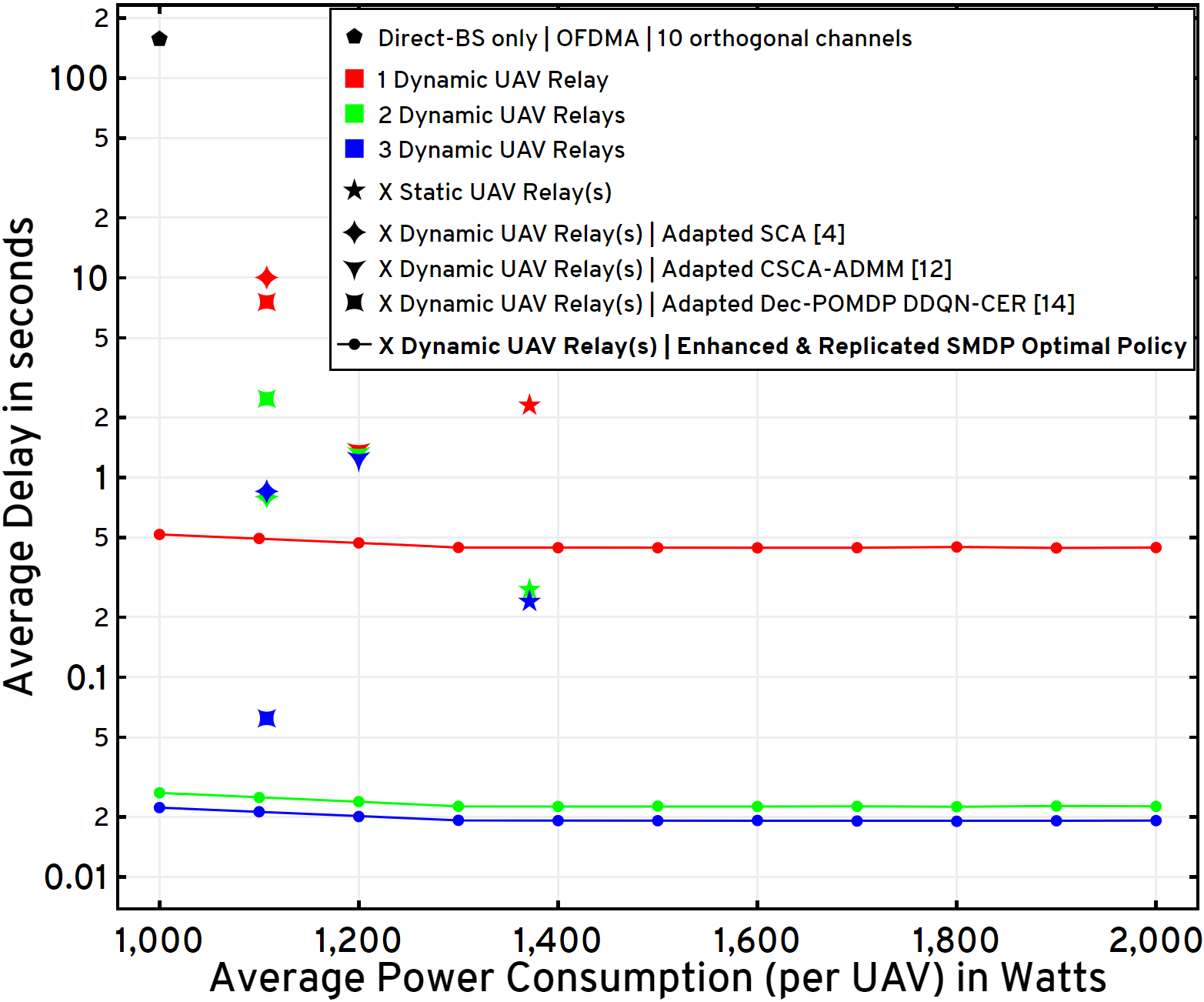}
         \vspace{-4mm}
         \label{F8}
     \end{subfigure}
     \vspace{-2mm}
     \caption{(a) Optimal waiting state policy for $L{=}1$ Mb with $P_{\mathrm{avg}}{=}1.2$ kW; (b) Average service latencies vs average power constraints for $L{=}1$ Mb.}
     \label{F7andF8}
     \vspace{-7mm}
\end{figure*}
\\\noindent{\textbf{Consensus-driven Conflict Resolution}}: When a new uplink transmission request originates in the cell, the UAVs already serving a GN continue to do so, i.e., they do no participate in the consensus-driven conflict resolution process. These relays, termed unavailable under this context, transition into their corresponding waiting states upon service completion. On the other hand, UAVs in the waiting states transition into their respective communication states. The BS along with these relays are deemed to be available. Studying the control frame exchanged by these available server nodes: the state flag is set to $1$, the GN position field is populated with the originating position of the request under consideration, and the cost-of-service field constitutes the delay-energy Lagrangian metric, i.e., $\frac{L}{\bar{R}_{GB}}$ for the BS and $\ell_{\nu}^{*}(\mathbf{s};\hat{r}_{U},1)$ for an available UAV. Upon sharing these metrics with each other, the available nodes arrive at a consensus on the best choice (delay-energy Lagrangian) for serving the request.

\vspace{-4mm}
\section{Numerical Evaluations}\label{V}
We use a channel bandwidth of $B{=}5$ MHz; for all links, NLoS attenuation constant $\kappa{=}0.2$, $1$-meter reference SNR $\frac{\beta_{0}P_{T}}{N_{0}B\Gamma}{=}40$ dB, LoS path-loss exponent $\alpha{=}2$, NLoS path-loss exponent $\tilde{\alpha}{=}2.8$, Rician $K$-factor parameters $k_{1}{=}1$ and $k_{2}{=}0.05$, and LoS probability parameters $z_{1}{=}9.61$ and $z_{2}{=}0.16$; UAV height $H_{U}{=}200$ m; BS antenna height $H_{B}{=}80$ m; maximum UAV velocity $V_{\mathrm{max}}{=}55$ m/s; and cell radius $a{=}1000$ m. Our UAV mobility power model uses the relationship and parameters detailed in \cite{SCA}. We solve an approximation of problem \eqref{eq:PolDecomp} by discretizing the SMDP state and action spaces, and applying value iteration; additionally, the optimal dual variable value is attained via projected sub-gradient ascent \cite{SubgradientMethods}. We discretize the states with $N_{\mathrm{sp}}{=}25$ equispaced radii values; similarly, $R_{\mathrm{sp}}{=}25$ corresponds to the equispaced radial velocity actions.\\
\indent{To} analyze the optimal waiting behavior of a UAV-relay in our control framework, we fix $P_{\mathrm{avg}}{=}1.2$ kW and $L{=}1$ Mb. Fig. \ref{F7andF8} shows that the UAV moves to and waits by flying around an optimal radius level (${\approx}95$ m) to address two considerations: to be well-positioned for future requests and to fly at the power-minimizing velocity so as to reduce its energy consumption. Moreover, we note that the angular velocity optimization process via exhaustive search adheres to the observations in \cite{SCA} about optimizing towards the minimum, i.e., $P_{\mathrm{min}}$ corresponding to $V{=}22$ m/s.\\
\indent{As} depicted in Fig. \ref{F7andF8}, for uplink transmission requests of size $L{=}1$ Mb from the $300$ GNs in the cell, we observe the following improvements in performance over custom network deployment heuristics and state-of-the-art frameworks, averaged over $10,000$ requests with a Poisson arrival rate of $1$ request every $60$ s. Considering only direct transmissions to a $10$-channel OFDMA BS at the cell center, we find a significant reduction in the average communication delay experienced by the GNs in the cell, by employing UAVs to relay data traffic. Also, we observe that employing dynamic UAVs with optimized trajectories result in lower service delays compared to static relay deployments---specifically, for a swarm of $3$ UAV-relays, with $L{=}1$ Mb and a per-UAV power consumption of $1.37$ kW, our solution services uplink transmission requests from the GNs $11{\times}$ faster than a static deployment of $3$ UAVs positioned equidistant from the cell center. Furthermore, with CVXPY implementations of joint multi-agent SCA strategies \cite{SCA, CSCA-ADMM} involving split conic solvers with $10^{6}$ iterations and an accuracy of $10^{-6}$, we note that our control system with $1$ UAV-relay exceeds the QoS performance offered by $3$ relays under these SCA approaches. Finally, with an average per-UAV power constraint of $1.1$ kW, our solution demonstrates $2{\times}$ faster service times relative to DDQNs \cite{DDQN}.

\vspace{-4mm}
\section{Conclusion}\label{VI}
In this paper, we detail the development of an adaptive framework for the decentralized orchestration of a swarm of rotary-wing UAV-relays in next-generation non-terrestrial networks. First, employing rate adaptation to leverage A2G channel dynamics, we specialize our system model to single UAV-relay settings and design the optimal request scheduling and trajectory optimization policy under an SMDP formulation (via value iteration and CSO). Next, we extend this single relay policy to distributed deployments of two or more UAVs by supplementing it with multi-UAV coordination heuristics and replicating it across the swarm. Numerical evaluations demonstrate that our solution delivers significant performance improvements over BS-only strategies, static UAV deployments, SCA approaches, and DDQN frameworks.

\vspace{-4mm}
\bibliographystyle{IEEEtran}
\bibliography{IEEEabrv,main}

\end{document}